\documentclass[%
preprint,
 amsmath,amssymb,
 aps, physrev,
]{revtex4-2}

\usepackage{graphicx}
\usepackage[normalem]{ulem}
\usepackage{dcolumn}
\usepackage{bm}
\usepackage{xcolor}
\usepackage{comment}

\begin{document}


\title{\textbf{Difference in Neoclassical Edge Flows Between Strongly Negative and Positive Triangularities in the XGC Gyrokinetic Simulation} 
}%

\author{S. Ku}
\email{Contact author: sku@pppl.gov}
\author{C.S. Chang}%
\author{R. Hager}
\affiliation{%
 Princeton Plasma Physics Laboratory, Princeton, NJ, USA
}%

\author{L. W. Schmitz}
\affiliation{
University of California, Los Angeles, CA, USA
}%
\author{A. O.  Nelson}
\affiliation{%
Columbia University, New York, NY, USA
}%

\date{\today}

\begin{abstract}

The neoclassical baseline study of a strongly negative triangularity (NT) plasma and the corresponding positive triangularity plasma is performed using the edge-specialized, total-f gyrokinetic code XGC. A DIII-D-like plasma is used, based on the negative triangularity discharge of DIII-D \#193793. An artificial positive triangularity (PT) equilibrium has been constructed to compare the edge rotation physics at the same triangularity strength, but with opposite sign, while keeping the same elongation and other geometric parameters. 
Carbon(+6) ions are added to the deuterium plasma at an experimentally relevant level.  By using the experimental profile of carbon toroidal rotation profile as an input, XGC finds that the deuteron rotation is significantly different from the carbon rotation at the inboard and outboard midplanes, mostly caused by the difference in the Pfirsch-Schl\"{u}ter rotation.  More importantly, significant difference in the X-point orbit loss physics, thus the rotation source, is found between the positive and negative triangularity equilibrium models. However, it is also found that the agreement between the present neoclassical simulation and the experimental NT data is validated only within the middle of pedestal slope, indicating the importance of edge turbulence.
This study could establish baseline for the multiphysics, multiscale studies that include turbulence of negative triangularity plasmas.  
\end{abstract}

\maketitle


\section{\label{sec:level1} Introduction}

Magnetic fusion energy research explores plasma equilibrium parameters to optimize confinement and performance. Triangularity, a plasma shape parameter defined as the difference between the geometric axis and the magnetic axis normalized by the minor radius, has emerged as a significant design variable with profound implications for plasma profiles and transport behavior\cite{Marinoni2009,Kikuchi2019, Marinoni2021,Austin2019,Coda2021,Happel2022, merlo2019, merlo2023negative, Wilson_2025, Balestri_2024, Faitsch_2018, Mariani_2024}. 

Recent investigations into negative triangularity have revealed intriguing plasma characteristics \cite{Marinoni2021NF,Nelson2022,Nelson2023,Thome2024, Wilson2025}.
Absence or reductions of edge localized modes (ELMs), which represents a particularly notable advancement, as these instabilities are known to cause significant heat and particle flux disruptions in traditional tokamak operations \cite{Nelson2023,Nelson_2024}.
Another nature of negative triangularity is its impact on edge transport barrier. Notably, achieving the H-mode, a high-confinement regime with the edge transport barrier, becomes practically impossible under conditions of sufficiently strong negative triangularity\cite{Nelson2023}. This unique behavior necessitates a comprehensive investigation into the fundamental mechanisms governing edge plasma properties under such unconventional geometric configurations.
It is important to systematically explore the intricate relationship between negative triangularity and edge plasma characteristics, seeking to elucidate the underlying physics differences between the strongly negative and positive triangularity plasmas. 

Most of the existing gyrokinetic simulation studies of negative triangularity plasmas have focused on core regions \cite{Marinoni2019, Merlo2015, Merlo2023,Li2024,Balestri_2024,Hoffmann_2025} or core-edge regions with limiter geometries and without carbon species \cite{Bernard2024}. In this work, we present a comprehensive neoclassical gyrokinetic study of diverted edge plasmas of DIII-D\cite{Luxon2002} to compare the base-line neoclassical differences between strongly negative and positive triangularity configurations using the edge-optimized, X-point included Gyrokinetic Code (XGC).  It is found that the most significant difference is in the edge plasma flow, caused by the X-point ion orbit loss phenomenon \cite{Chang2002,Ku2004, deGrassie2015}.

We note here that the present study is to establish the neoclassical baseline difference between negative and positive triangularity equilibrium and not intended to compare with experimental observations since the turbulence physics is omitted. We also emphasize that the manufactured positive triangularity equilibrium and the divertor geometry presented here are not something DIII-D can create experimentally. Even if the positive triangularity shape could be achieved experimentally, it would likely access H-mode or exhibit different plasma profiles due to altered transport coefficient. 

This paper is organized as follows: Sec.~\ref{sec.method} introduces the equilibrium profiles with experimental negative triangularity and manufactured positive triangularity, and presents the gyrokinetic simulation model. Sec.~\ref{sec.orbitloss} illustrates the difference in the X-point ion orbit loss physics between positive and negative triangularity equilibrium. Sec.~\ref{sec.neoclassical} presents the neoclassical gyrokinetic simulations and the results. Conclusion and discussions are presented in Sec. \ref{sec.conclusion}.

\section{Equilibrium Profiles and Simulation Method\label{sec.method}}


\begin{figure}
    \centering
    \includegraphics[width=0.93\textwidth]{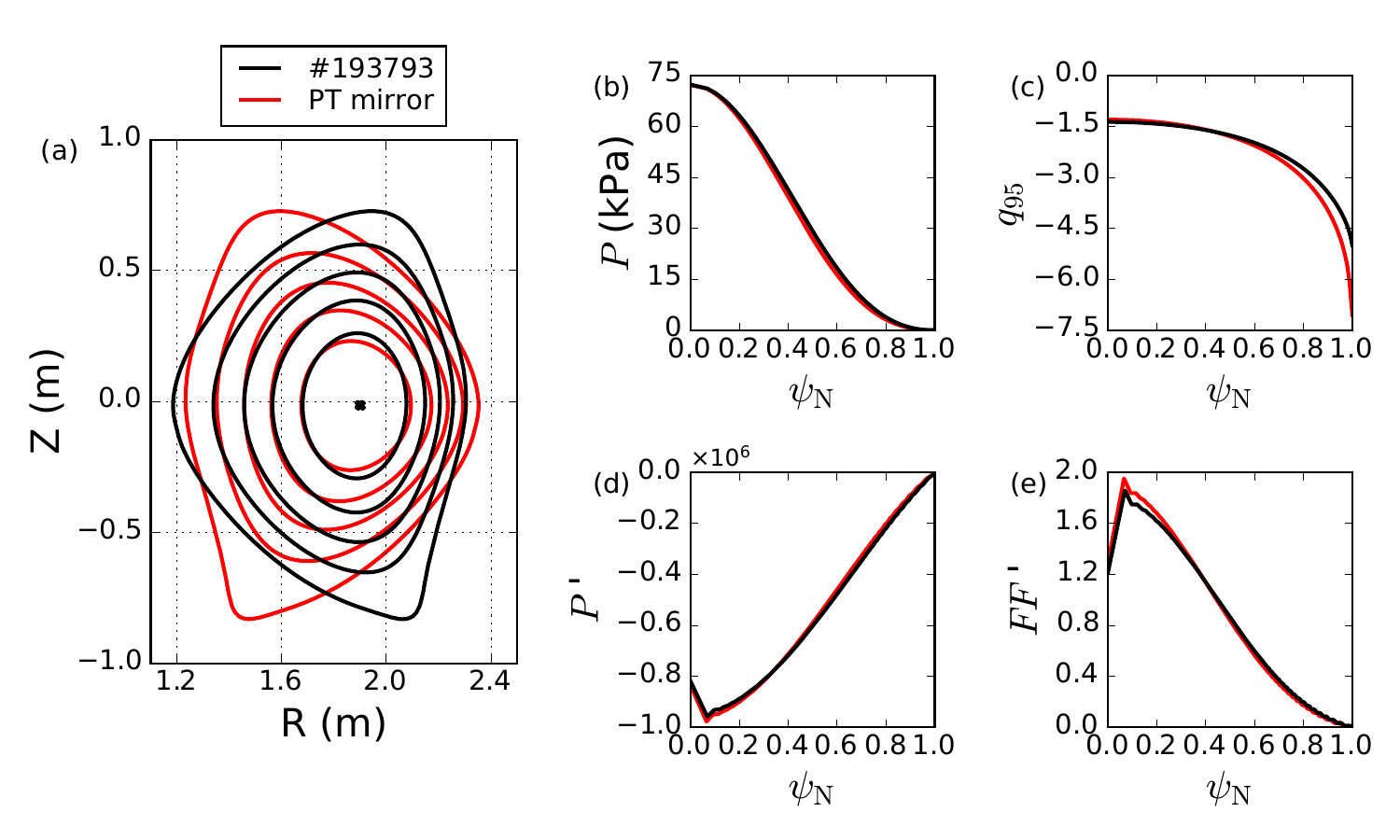}
    \caption{ (a) poloidal flux, (b) pressure, (c) safety factor, (d) pressure gradient (P'), and (e) the product of poloidal current function and its derivative (FF') from Grad-Shafranov equation of  NT (black) and PT (red).}
    \label{fig:geometry}
\end{figure}

XGC (X-point Gyrokinetic Code), is a multiscale 5-dimensional gyrokinetic code that solves the turbulence-neoclassical-neutral transport physics together, and specialized for tokamak edge plasma based on the particle-in-cell technique combined with a velocity-space grid method\cite{Ku2016}. The gyrokinetic equations, the field equations, and the numerical algorithms are detailed in section II and III of reference \cite{Ku2018}. Present neoclassical simulations are achieved by restricting the field solver to axisymmetric electrostatic potentials, but accounting for radial and poloidal variations.

We base our plasma models on the DIII-D strongly negative triangularity discharge \#193793 at 1,700 ms, which features the following parameters:
the upper triangularity $\delta^{u}=-0.35$, the lower triangularity $\delta^{l}=-0.56$, plasma current $I_p$ =0.59 MA, toroidal magnetic field at the magnetic axis $B_T$=1.4 T, and the safety factor $q_{95}\simeq4.8$. 
The magnetic geometry is shown as the black lines of Fig~{\ref{fig:geometry}} (a).
In this configuration, the toroidal magnetic field, the plasma current, and the parallel flow all circulate counter-clockwise when viewed from above the tokamak, resulting in the downward-directed $\nabla B$-drift. 

\begin{figure}
    \centering
    \includegraphics[width=0.53\textwidth]{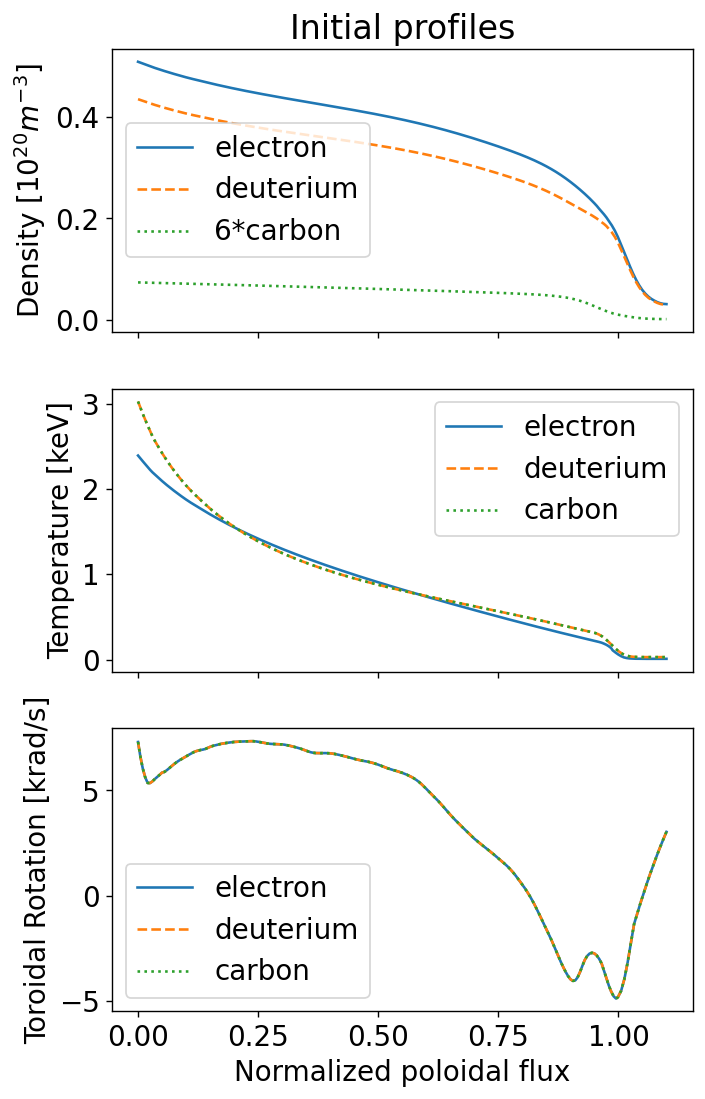}
    \caption{ Initial profiles of density (top), temperature (middle), and toroidal rotation (bottom). Carbon density is multiplied by the charge number, 6. Initial temperature of deuterium and carbon are identical. The toroidal rotations are identical for all species.}
    \label{fig.profiles}
\end{figure}

Using this NT equilibrium, we create a corresponding positive triangularity (PT) equilibrium by mirroring the separatrix and limiter/divertor shapes. 
The flux-surface-averaged plasma profiles, including the safety factor (q) profile, in PT are matched as closely as possible to those from NT, allowing for the comparison without concerns arising from the plasma profile differences.  
This careful matching of equilibrium parameters ensures that any observed differences in plasma behavior can be attributed specifically to the change in triangularity rather than to variations in other plasma parameters.
%


Since DIII-D divertor/wall geometry does not allow such a ``NT-matching '' PT equilibrium, the wall geometry is simplified to enable a more proper theoretical comparison between NT and PT.


Our simulations include kinetic electrons, deuterium ions, and fully ionized carbon ions (charge number 6). The initial plasma particle distributions are modeled as local Maxwellians with density, temperature, and toroidal rotation profiles as shown in Figure~\ref{fig.profiles}.
These initial profiles are derived from experimental measurements, with the deuterium temperature initially assumed to match the carbon temperature. The toroidal rotation profile is calibrated to ensure that the simulated carbon rotation at the end of the simulation closely corresponds to experimental carbon rotation measurements at the low-field side midplane. This calibration process is detailed in section \ref{sec.neoclassical}.
Initially, electron and deuterium rotation profiles are set to be identical to the carbon rotation profile and let to evolve according to total-f gyrokinetic equations \cite{Ku2016} under the given magnetic and wall geometries. 
%
The simulations incorporate a fully nonlinear Fokker-Planck collision operator to allow possible non-Maxwellian particle distribution around the magnetic separatrix and in the scrape-off-layer (SOL)\cite{Hager2016}. 
Neutral atomic physics is also modeled using a Monte-Carlo approach for neutral particles \cite{Stotler2017}. 

Simulations are performed on the Perlmutter supercomputer at the National Energy Research Scientific Computing Center (NERSC).
For neoclassical simulations, we utilized 8 CPU nodes for 24 hours with 210 million marker particles for each species distributed across a triangular mesh of 24,000 vertices, resulting in approximately 9,000 particles per mesh vertex. The velocity space resolution used for  Coulomb collisions, neutral atomic physics, and particle noise reduction\cite{Ku2016} consisted of a 33 $\times$ 31 grid (perpendicular velocity $\times$ parallel velocity).  

\section{Discussion on the X-point ion orbit loss \label{sec.orbitloss}}

First, we discuss the difference in the X-point orbit-loss physics \cite{Chang2002,Ku2004, deGrassie2015} between NT and PT plasmas to help the readers understand the simulation results presented in the next section. We limit the ion X-point orbit loss discussion in the absence of electric field; meaning ``test particle simulations.'' We note here that both the X-point orbit loss effect and resulting toroidal flow are exaggerated in these test particle simulations since they lack the mitigating effect of the radial electric field. 

\begin{figure}
    \centering
    \includegraphics[width=0.93\textwidth]{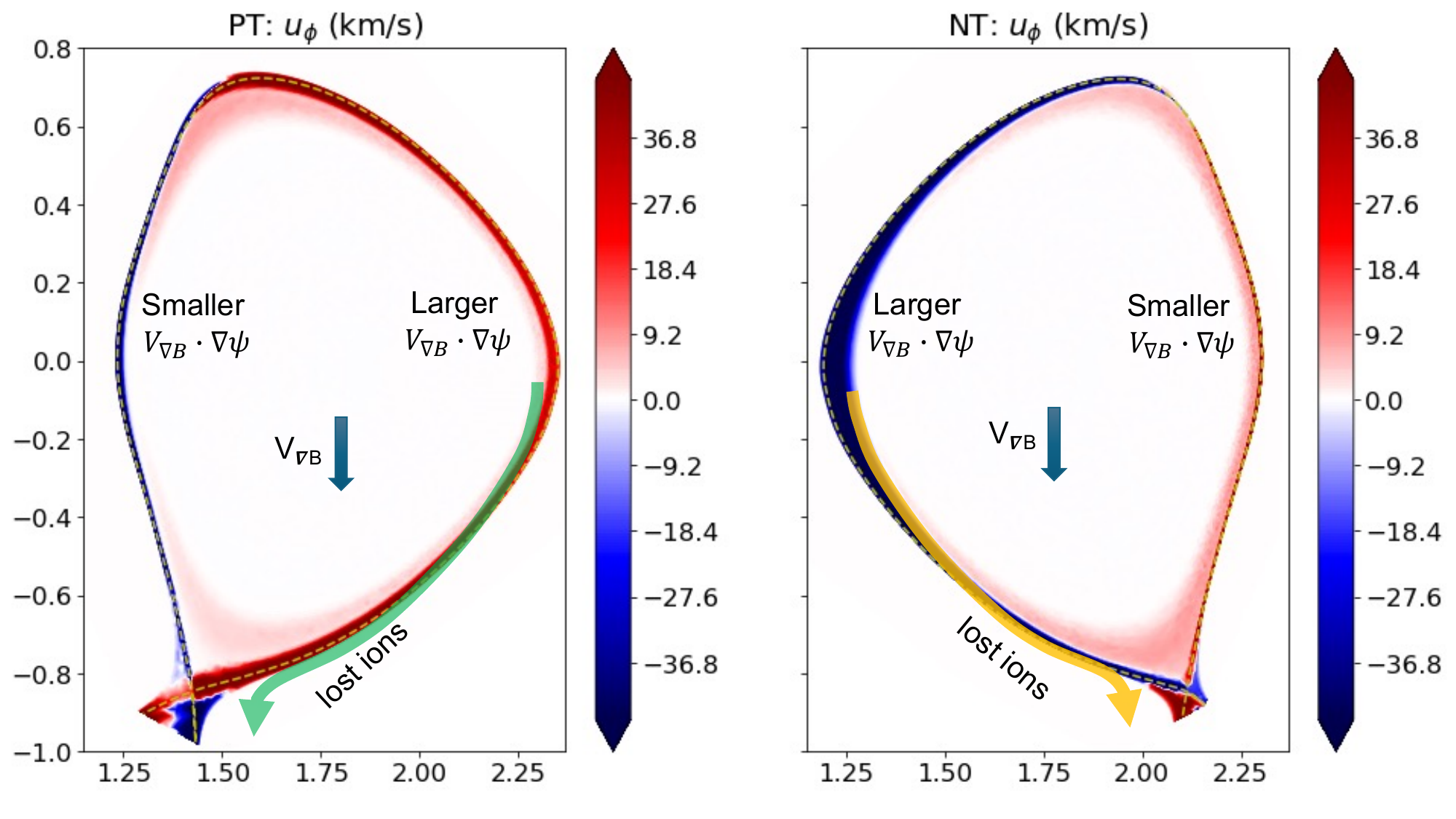}
    \caption{ Toroidal flow ($u_\phi$) structure from test particle simulations of PT (left) and NT (right) plasmas without electric field. The color represents the direction and magnitude of toroidal flow, with red indicating positive toroidal (co-current) flow and blue indicating negative toroidal (counter-current) flow. The far scrape-off-layer is not displayed for better visualization. The curved arrows illustrate cartoon picture of ion particle trajectories projected to the poloidal planes accounting for the parallel motion and the $\nabla B$ drift ($V_{\nabla B}$). Particles with opposite parallel direction to the direction of cartoon trajectory on the side of strong flux surface curvature have a higher probability of being lost. The magnitude of the curvature determines how large the radial component of the $\nabla B$ drift is on the above or below the midplane. $\psi$ is the poloidal flux.
    }
    \label{fig.test}
\end{figure}

For this test particle study, we have used a stationary Maxwellian distribution with flat density ($2 \times 10^{19} m^{-3}$) and temperature (2 keV) profiles for the initial conditions. The marker particles are distributed throughout the entire volume from the magnetic axis to the wall.
Since we use flat density and temperature profiles, no Pfirsch-Schl\"{u}ter (P-S) flow is generated in these test particle simulations.
The details of the P-S flow will be discussed in Sec.~\ref{sec.neoclassical}. We note that the flow in these test particle simulations arise solely from the X-point orbit loss physics. Collisions and neutral ionization processes are disabled. Figure~\ref{fig.test} illustrates the two-dimensional toroidal flow structure that develops as a consequence of X-point ion orbit loss.

The cartoon picture in Fig.~\ref{fig.test}, depicts the ion orbits (green arrow for PT and yellow arrow for NT) more prone to  X-point orbit loss due to the larger $\nabla B$ drift ($V_{\nabla B}$) in the greater plasma volume area. This asymmetry results in a net positive (negative) co-current toroidal flow in the PT (NT) plasma.
%
Consequently, the X-point ion orbit-loss has tendency to generate a counter-current (co-current) torque at the plasma edge in NT (PT) plasma.

\begin{figure}
    \centering
    \includegraphics[width=0.52\textwidth]{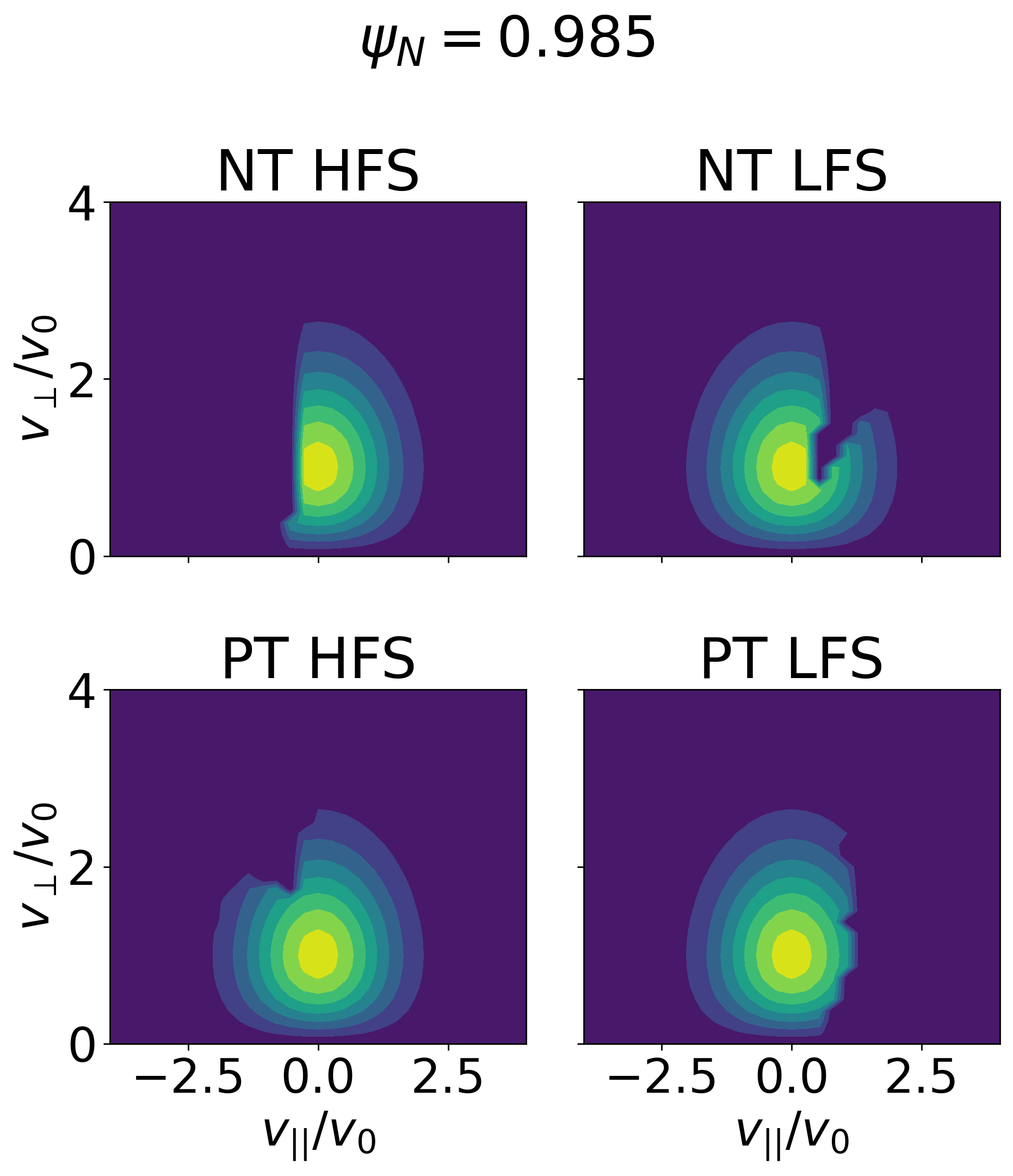}
    \caption{Velocity space hole from the test particles simulations without radial E-field. Top (bottom) graphs are from negative (positive) triangularity. Left (right) graphs are measured at high (low) field side midplane with $\psi_N$ = 0.985. $v_0$ is thermal velocity with flat 2 keV profile, which is for better visualization. The wiggles are from particle noises and limited resolution of velocity space grid, 33 by 31}
\label{fig.v-hole}
\end{figure}

Figure~\ref{fig.v-hole} shows the velocity space hole from the test particle simulations at the normalized poloidal flux $\psi_N$=0.985. The particle distribution functions at the midplanes confirm greater X-point loss-hole size at the strong curvature side: the high-field-side (HFS) for NT and the low-field-side (LFS) for PT. 
The present figures are also consistent with the previous report\cite{Nishimura2020, Kramer2025} that the velocity space hole of orbit loss is greater in an NT equilibrium.

\begin{figure}
    \centering
    \includegraphics[width=0.48\textwidth]{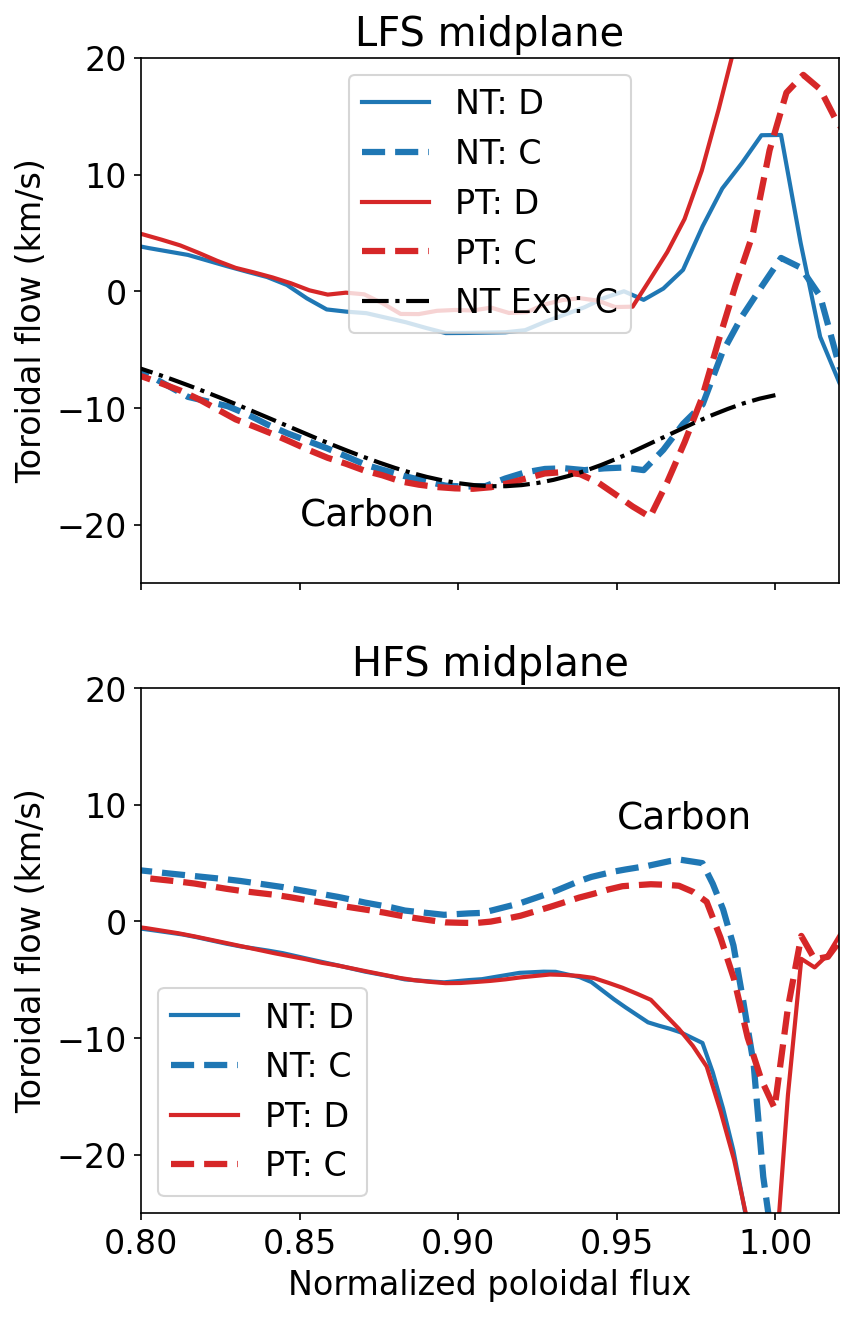}
    \caption{Toroidal flow of deuterium (solid lines) and carbon impurities (dashed lines) at the low-field-side midplane (top) and high-field-side midplane (bottom) from neoclassical simulations of PT (blue) and NT (orange) geometries. Black dash dot line in the top figure represent experimental measurements of carbon impurities.}
    \label{fig.flow_neo}
\end{figure}

\section{Neoclassical flows and radial electric field\label{sec.neoclassical}}

We now perform the total-f XGC neoclassical simulations with self-consistent $n=0$ electric field solutions.   
Reference \citenum{Chang2002} reported that, in a realistic scenario, the radial electric field is generated by X-point ion orbit loss, which is then largely mitigated by the neoclassical return current and orbit squeezing due to the $E \times B$ shearing action to leave only enough ion loss to keep up with the electron loss. 
In these self-consistent total-f simulations, we include carbon impurities with a charge number 6, fully nonlinear Fokker-Planck Coulomb collisions, and Monte Carlo neutral particles with ionization, charge exchange and wall-recycling. Our simulations reached a neoclassical quasi-steady state after 0.36 ms, which is longer than the ion-ion collision time, $\tau_{ii}\simeq $ 0.13 ms in the middle of the pedestal.
We reiterate that the X-point ion loss and the generated flow in the test particle simulations of Sec.~\ref{sec.orbitloss} are exaggerated due to the absence of radial electric field, and the self-consistent simulations in this section provide the realistic magnitude of X-point orbit loss effects.

Figure \ref{fig.flow_neo} shows the toroidal flow profiles of deuterium and carbon species at the low-field-side (top) midplane and the high-field-side (bottom) midplane at t=0.36 ms.
As observed in test particle simulations, PT exhibits a significantly higher co-current edge flow (at $\psi_N > 0.98$) for the main ion species at the low-field-side midplane compared to NT. By comparing with the high-field side toroidal flows (bottom figure), it can also be seen that the flux-surface-averaged edge toroidal flow in PT is also higher than in NT.

\begin{figure}
    \centering
    \includegraphics[width=0.48\textwidth]{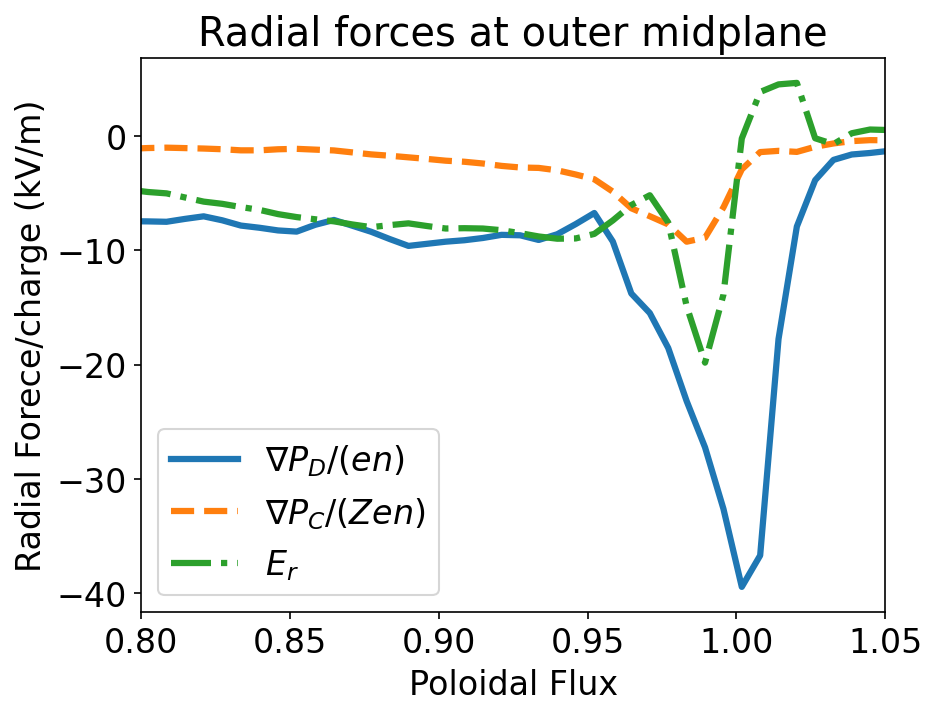}
    \caption{ Radial force per unit charge on the low field side midplane from neoclassical simulations of NT plasma at t=0.36 ms. The blue solid line is the radial force by deuterium pressure ($\nabla P_D/en_D)$, the orange dashed line is the radial force by carbon pressure ($\nabla P_C/(Z en_C)$), and the green dash-dotted line is the minus radial electric field ($E_r$), where $P_D$ and $P_C$ are deuterium and carbon pressure, $n_D$ and $n_C$ are deuterium and carbon density, and $Z$ is the charge number of carbon, 6.}
    \label{fig.rad_forece}
\end{figure}

One notable observation is that the toroidal rotation profiles at the midplanes differ significantly between the deuterium and the carbon species for both NT and PT.  We find that this is due to the difference in P-S flows. The P-S flow arises due to the incompressibility condition of plasmas flows, and expressed as
\begin{equation}
\label{eq.ps_flow}
    {\vec V}_{PS} =  \left( \frac{1}{Zen} \frac{dP}{dr} - E_r \right) \frac{dr}{d\psi} h \vec B,
\end{equation}
where $e$ is the elementary charge, $Z$ is the charge number, $n$ is the density, $P$ is the pressure, and $E_r$ is the radial electric field, $\vec B$ is the magnetic field vector, and h is a geometrical factor \cite{Nemov1990,Kumar2017}, which is given by
\begin{equation*}
\vec B \cdot \nabla h = - 2 \frac{(\vec B \times \nabla B)\cdot \nabla \psi}{B^3}, ~ \left< h B^2 \right> = 0
\end{equation*}
It can be easily seen from Eq.~\ref{eq.ps_flow} that the Z-dependence in the pressure gradient force makes the P-S flow speed of carbon species to differ from that of deuteron species.  As a matter of fact, for the D-species in the core region $\psi_N < 0.95$, Fig. \ref{fig.rad_forece} shows that the pressure term and the $E_r$ term nearly cancel each other, which is not true for the carbon species.  In the pedestal $\psi_N > 0.95$, the cancellation in the D-species between the pressure term and the $E_r$ term disappears.

Figure \ref{fig.rad_forece} shows the radial force per unit charge in Eq.~\ref{eq.ps_flow} on the low field side midplane from neoclassical simulations of NT plasma. In the region of $\psi_N < 0.95$, the deuteron pressure forces and the radial electric field force nearly cancel each other. This gives smaller net component of $\frac{dP}{dr}/(Zen) - E_r $ in Eq.~\ref{eq.ps_flow} for deuteron species. However, this cancellation is weak for the carbon species since carbons have charge number 6 and the radial E-field term in the P-S flow becomes dominant. 

\begin{figure}
    \centering
    \includegraphics[width=0.48\textwidth]{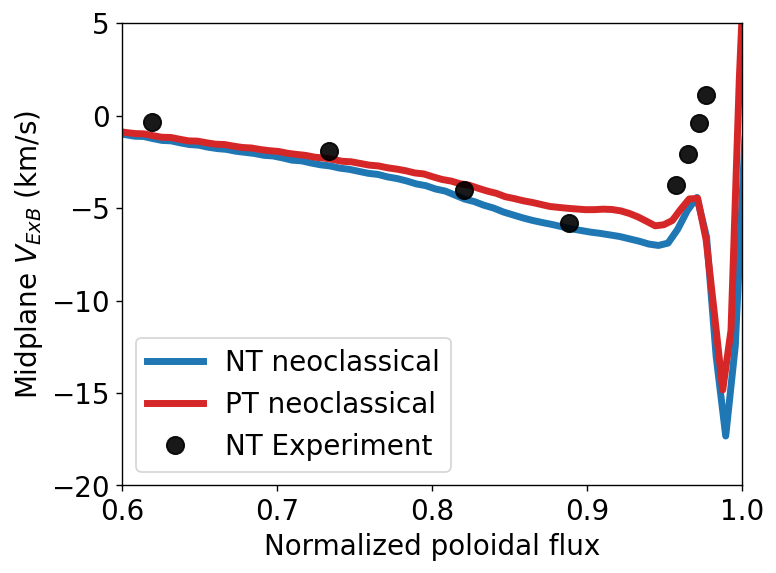}
    \caption{ $ V_{E \times B}$ on the low field side midplane from neoclassical simulations, time averaged between t=0.33 ms and t=0.36 ms. Blue and read lines represent NT and PT, respectively. The black dots represent the experimental measurements.}
    \label{fig.vexb_neo}
\end{figure}

Upper plot in Figure \ref{fig.flow_neo} shows together the experimentally measured carbon flow profile in the NT edge (black dash-dot line), measured using Doppler backscattering diagnostics.  It can be seen that the neoclassical simulation (the blue dash line) reproduces the experimental carbon rotation profile (the black dash-dot line) up to $\psi_N < 0.98$.  Beyond that radial location into the separatrix and SOL, the present neoclassical simulation result significantly deviates from the experimental observation.  This disagreement can also be seen in the radial electric field comparison. In Fig. \ref{fig.vexb_neo}, we show the $V_{E \times B}$ on the low field side midplane from neoclassical simulations and from experiment.  Similar disagreement can be seen at $\psi_N > 0.98$, indicating that the neoclassical physics may not be enough in predicting the plasma behavior in the far edge of tokamak NT plasmas. Within the ``more confident'' radial region ($\psi_N < 0.98$), we do not find a significant difference in the toroidal or ExB rotation profiles between PT and NT plasmas. In our future studies, the turbulence feature will be turned on in XGC -- which requires much more computational resources -- and the result will be reported.



\section{Conclusion\label{sec.conclusion}}

In the negative triangularity edge research, understanding the baseline neoclassical property needs to be established for a better investigation of turbulence property and its impact on the edge confinement. We have performed the neoclassical study of the plasma flow difference in the edge of strongly negative and positive triangularity plasmas by using an experimental negative triangularity plasma equilibrium and a manufactured but equivalent otherwise, positive triangularity equilibrium.  The total-f gyrokinetic code XGC has been used.

It is found that 1) toroidal flow speed of the carbon is significantly different from that of main ions in the edge mostly from the Pfirsch-Schluter effect, 2) the X-point ion orbit loss physics makes the PT plasma flow to be shifted more into the positive direction compared to the NT plasma both in the outer and inner midplanes in the far edge $\psi_N > 0.98$, 3) however, experimental comparison indicates that the plasma rotation property in the far edge ($\psi_N > 0.98$) is not well-predicted by neoclassical simulation and turbulence may need to be added to see how the far edge rotation and $Er$ profiles are modified, and 4) there is not much neoclassical difference at $\psi_N < 0.98$ between NT and PT plasmas.

It is the suggestion of the present study that the X-point ion orbit loss physics, Pfirsch-Schluter physics, and turbulence need to be considered self-consistently for a higher-fidelity study of the negative triangularity plasma.  Our subsequent report will be aimed for such a study.

\begin{acknowledgments}

This research is based upon work supported by U.S. Department of Energy, Office of Science, Office of Fusion Energy Sciences, via  PICSciE, Princeton University and the SciDAC-5 Center for Edge of Tokamak OPtimization (CETOP) under Awards DE-AC02-09CH11466, University of California Los Angeles under Awards DE-SC0020287, and Columbia University under Awards DE-SC0022270.

This material is based upon work supported by the U.S. Department of Energy, Office of Science, Office of Fusion Energy Sciences, using the DIII-D National Fusion Facility, a DOE Office of Science user facility, under Award(s) DE-FC02-04ER54698.

This research used resources of the National Energy Research Scientific Computing Center (NERSC), a Department of Energy Office of Science User Facility using NERSC award FES-ERCAP0032875.

\end{acknowledgments}

\section*{Disclaimer}
This report was prepared as an account of work sponsored by an agency of the United States Government. Neither the United States Government nor any agency thereof, nor any of their employees, makes any warranty, express or implied, or assumes any legal liability or responsibility for the accuracy, completeness, or usefulness of any information, apparatus, product, or process disclosed, or represents that its use would not infringe privately owned rights. Reference herein to any specific commercial product, process, or service by trade name, trademark, manufacturer, or otherwise does not necessarily constitute or imply its endorsement, recommendation, or favoring by the United States Government or any agency thereof. The views and opinions of authors expressed herein do not necessarily state or reflect those of the United States Government or any agency thereof.

Raw data were generated at the National Energy Research Scientific Computing Center
large scale facility. Derived data supporting
the findings of this study are available from the
corresponding author upon reasonable request.

\bibliography{NT_XGC}

\end{document}